\documentclass[preprint,superscriptaddress,showpacs,preprintnumbers,amsmath,amssymb]{revtex4}
\usepackage{graphicx}

\begin{document}

\title{Partial Clustering in Binary Two-Dimensional Colloidal Suspensions}

\author{Norman~Hoffmann}
\affiliation{%
Institut f\"ur Theoretische Physik II,
Heinrich-Heine-Universit\"at
D\"usseldorf, Universit\"atsstra\ss e 1, D-40225 D\"usseldorf, Germany\\
}%
\author{Florian~Ebert}
\affiliation{%
Fachbereich f\"ur Physik, Universit\"at Konstanz, 
D-78457 Konstanz, Germany
}%
\author{Christos~N.~Likos}
\affiliation{%
Institut f\"ur Theoretische Physik II,
Heinrich-Heine-Universit\"at
D\"usseldorf, Universit\"atsstra\ss e 1, D-40225 D\"usseldorf, Germany\\
}%
\author{Hartmut~L\"owen}
\affiliation{%
Institut f\"ur Theoretische Physik II,
Heinrich-Heine-Universit\"at
D\"usseldorf, Universit\"atsstra\ss e 1, D-40225 D\"usseldorf, Germany\\
}%
\author{Georg~Maret}
\affiliation{%
Fachbereich f\"ur Physik, Universit\"at Konstanz, 
D-78457 Konstanz, Germany
}%

\date{\today}

\begin{abstract}
Strongly interacting binary mixtures of superparamagnetic
colloidal particles confined
to a two-dimensional water-air interface are examined  by
theory, computer simulation and experiment. The mixture exhibits a partial
clustering in equilibrium: in the voids of the  matrix of
unclustered big particles, the small particles
form subclusters with a sponge-like topology which is accompanied by a
characteristic small-wave vector
peak in the small-small structure factor. This partial clustering is a
general phenomenon occurring for strongly coupled negatively
non-additive mixtures. 
\end{abstract}

\pacs{61.20.-p, 64.75.+g, 82.70.Dd, 61.20.Ja}

\maketitle

Self-organization processes in suspensions of colloidal particles
are of key relevance for the construction of nanodevices
with desired mechanical, rheological
  and optical properties. In particular,
colloidal aggregates or clusters can be used as
supramolecular building blocks, motivated by the quest for 
photonic band-gap materials \cite{Pine}, molecular-sieves
\cite{Kecht,Lutz} and micro-filters  with desired
porosity \cite{Goedel}. Colloidal cluster formation is mainly
controlled by the effective interparticle forces \cite{Pusey}: strong 
attractive forces
between the colloidal particles will drive them
towards irreversible coagulation into dense aggregates while
repulsive forces in a stable suspension keep the particles apart, avoiding
clustering. More recently, it was predicted that
a combination of short-ranged attractive
and long-ranged repulsive forces can lead to equilibrium clusters
\cite{Sear,Imperio,Malescio} for the case of hard colloids, whereas
particular classes of ultrasoft colloids can show clustering even
in the complete absence of attraction \cite{mladek:prl:06}.
The cluster morphology critically depends
on the details of the interaction, being, e.g., spherical
or worm-like with an internal spiral structure \cite{sciortino_mossa,mossa_co}.
The equilibrium clustering has been confirmed in recent
experiments \cite{Stradner,Campbell,Weitz} of one-component
weakly charged suspensions with nonadsorbing polymer
additives.

In this Letter, we address equilibrium clustering in {\it two-component}
suspensions of big and small colloidal particles. Our motivation 
is twofold: first, any realistic sample consists of several components
and hence considering a binary mixture is an important step towards
a control of novel composite materials.
Second, and more fundamentally, mixtures  exhibit much richer behavior than
their one-component counterparts in general. Therefore, there is need to 
explore
whether and how the scenario of equilibrium clustering occurs
in two-component systems. Specifically, we consider here
a binary  system of superparamagnetic colloidal particles
that are confined to a planar water-air interface and
exposed to an external magnetic field perpendicular
to the interface. The magnetic field induces a magnetic dipole moment
on the particles, resulting into an effective repulsion
between all parallel oriented dipole moments, which scales with the inverse
cube of the particle distance. This enables a direct comparison between
theories
and computer simulations based on a pairwise dipole-dipole interaction
potential and renders these suspensions 
into an ideal model system \cite{Maret_1},
allowing thereby to additionally confront the theoretical findings
with direct experimental data. 
The advantage of the availability of
experimental systems has been exploited to study two-dimensional
melting
in the one-component system \cite{Maret_2,Maret_3}
and for the study of the 
dynamics \cite{Naegele} and the glass transition \cite{Koenig}
of the binary system. Yet, cluster formation 
for this system has not been studied to-date. The system at hand
presents a host of novel features: it is a two-dimensional
system,
it has clearly defined interactions
between all its constituent particles, which take a simple form
and are steered by a single and readily tunable external field;
it is experimentally realizable and allows direct comparison 
with theory; and, finally, it is apt to direct visual observation.

We find that colloidal clustering is specific to the components
in the repulsive mixture: subclustering is possible where only the
small particles form clusters but the big ones are unclustered, despite
of the fact that 
they are strongly interacting with the small ones. 
This {\it partial clustering} 
does not occur in the effective one-component colloid systems 
since the structure of the additives simply follows that of the colloids.
We trace the origins of the phenomenon to the {\it negative 
nonadditivity} \cite{roth:evans:epl:01} of the mixture, 
which expresses the fact that the cross-interaction between
unlike species is less repulsive than the sum of the interactions
between like species. Quantitatively, negative nonadditivity, $\Delta < 0$, 
is defined as follows. Let $v_{ij}(r)$, $i,j = 1,2$
be the three interaction potentials between all species and 
introduce the corresponding Barker-Henderson effective hard core
diameters $\sigma_{ij} = 
\int_0^{\infty}{\rm d}r\{1 - \exp[-v_{ij}(r)/k_{\rm B}T]\}$, 
with Boltzmann's constant $k_{\rm B}$ and the absolute temperature $T$.
Then, $\Delta = 2\sigma_{12} - (\sigma_{11} + \sigma_{22})$.
The partial clustering effect
should therefore be more generic to negatively nonadditive binary mixtures of,
e.g., charged colloids \cite{Klein}
or star-polymers and linear chains \cite{cnl:epl:05}.
However, partial clustering is 
absent in strictly additive ($\Delta = 0$) hard sphere mixtures
and in 
colloid-polymer mixtures, which are positively-nonadditive ($\Delta > 0$).
The structural signature of this phenomenon is the appearance
of an additional
small-wave-vector pre-peak in the small-small structure factor
correlations, which
does not show up in the big-big structure
factor. Based on a topological
analysis using the Euler characteristic, we show that the 
small-particles 
have a nontrivial, sponge-like structure. We base our findings on
liquid-integral equation theory, computer simulations  
and on video microscopy data of real samples, finding very good
agreement between all.

\begin{figure}[t]
  \begin{center}
    \includegraphics[width=14cm,clip=true]
    {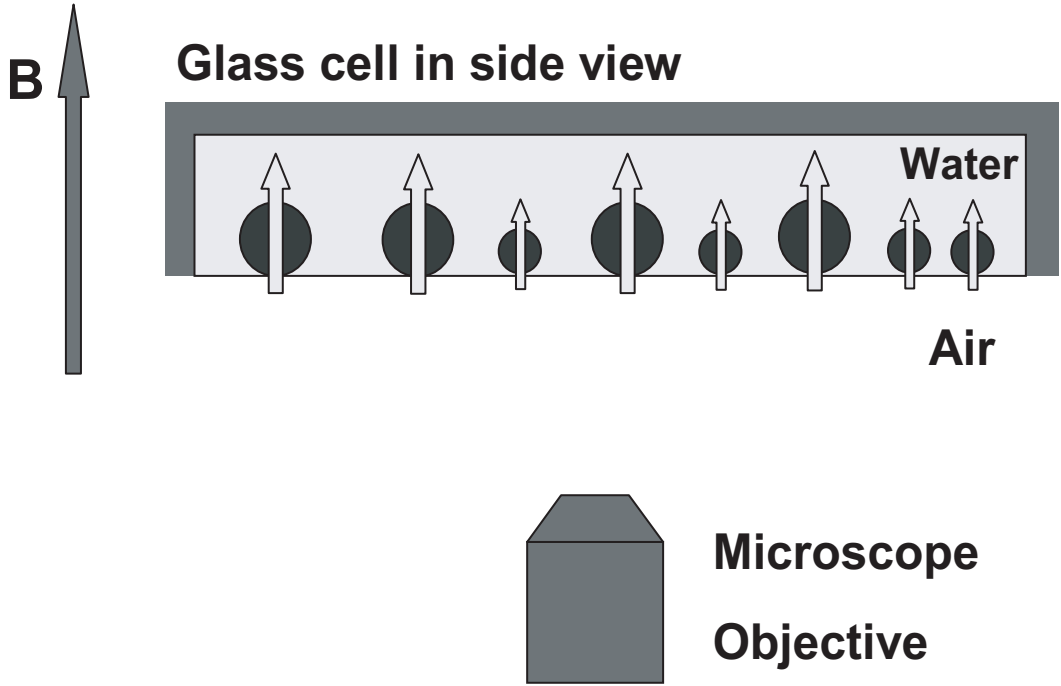}
    \caption{Schematic view of the setup: binary mixture of
    superparamagnetic colloidal 
    particles at an air-water interface in an external magnetic field
    ${\bf B}$ perpendicular to the plane.}
  \label{fig:setup}
  \end{center}
\end{figure}

In our theoretical model, we consider two different species of
colloidal particles moving freely in a two-dimensional plane, which
corresponds to the experiment (see Fig.\ \ref{fig:setup}). Each
component is characterized by its partial density $\rho_i$ and
its susceptibility $\chi_i$, $i=1,2$. 
The external magnetic field ${\bf B}$ standing perpendicular 
to the plane, induces in each
particle a magnetic moment $m_i=\chi_i B$, $i=1,2$. The particles are
superparamagnetic, i.e., the magnetic dipole of each
species perfectly aligns with the external field. We model all particles
as point-like but refer thereafter to the particles having the larger
susceptibility as the `big' (species 1) and those with
smaller susceptibility as `small' (species 2).
The particles interact via
purely repulsive dipole-dipole pair potentials $\beta v_{ij}(x) =
\Gamma_{ij}/x^{3}$, $i,j=1,2$, where $\beta = (k_{\rm B}T)^{-1}$ and
$x$ denotes the distance between any two
particles scaled over the average interparticle separation between 
{\it big} particles, $x\equiv r\sqrt{\rho_1}$.
The interaction strengths $\Gamma_{ij}$ are given as $\Gamma_{ij} =
\beta \chi_i \chi_j B^{2}/a_{11}^3$. 
The system is fully characterized by three
quantities: the density ratio $\rho_2/\rho_1$, the susceptibility ratio
$\chi_2/\chi_1 < 1$ and one of the three interaction strengths, which
we pick to be $\Gamma_{11}$ in what follows.

\begin{figure}[t]
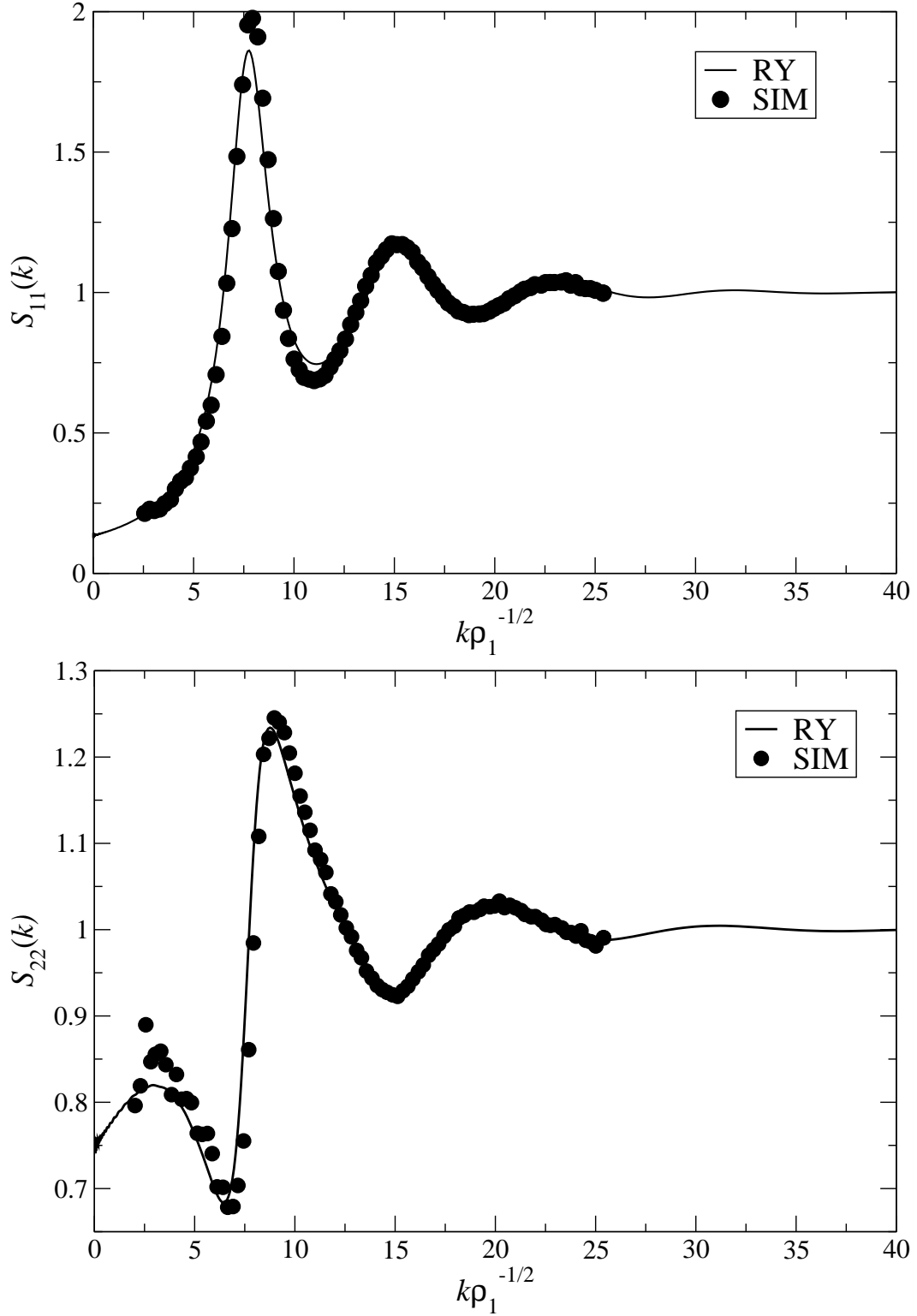

  \begin{center}
    \includegraphics[width=14.0cm,clip=true]
    {./fig2a.eps}
    \includegraphics[width=14.0cm,clip=true]
    {./fig2b.eps}
    \caption{Partial structure factors $S_{11}(k)$ for the big
    particles and $S_{22}(k)$ for the small particles. Computer
    simulation results
    are compared to the RY closure for the following parameter
    combination: $\Gamma_{11}=4.0$, $\rho_2/\rho_1=0.5$, $\chi_2/\chi_1=0.5$.}
  \label{fig:sofk_sim_theory}
  \end{center}
\end{figure}

We have performed Monte Carlo computer
simulations \cite{allen} in a square box with periodic boundary
conditions. Finite-size effects were carefully checked for by varying the 
box size from 500 to 2000 particles. Additionally,
we solved the binary Ornstein-Zernike relation \cite{Hansen} 
in two spatial dimensions, using the
Rogers-Young closure (RY) \cite{Rogers}. 
The RY gives reliable
results  in comparison to the simulation 
for all parameter
combinations investigated, see 
representative results in Fig.\ \ref{fig:sofk_sim_theory}. Thus,
we rely on this closure to calculate the
structure of the fluid. In Fig.\ \ref{fig:sofk_sim_theory},
the clustering of the small particles is manifest by an additional
small-wave-vector peak in the structure factor $S_{22}(k)$ whereas 
$S_{11}(k)$ is devoid of any such signature. Hence, the
formation of small particles subclusters in absence of big particle
clusters is witnessed by structural information in reciprocal space.

The experimental system is fully described by the theoretical
model introduced above and is explained in detail 
elsewhere \cite{Maret_1,Maret_2}. We use superparamagnetic 
particles \cite{DYNABEADS}
with
susceptibilities $\chi_1=6.2\times 10^{-11}\,{\rm Am{^2}/T}$ and 
$\chi_2=6.6\times 10^{-12}\,{\rm Am^{2}/T}$,
which are suspended in a free-hanging, flat
water droplet
attached to a top-sealed glass ring (diameter=$8\,{\rm mm}$, see 
Fig.\ \ref{fig:setup}).
Stabilization with Sodium Dodecyl Sulfate prevents the
particles from aggregation. Due to high mass density
($\rho_1=1.3\,{\rm kg/dm^{3}}$, $\rho_2=1.5\,{\rm kg/dm^{3}}$), both
types of particles are
pinned down to the water-air interface by gravity and form an ideal,
two-dimensional
monolayer of binary dipoles.
The relatively small gravitational lengths of $l_1=8\,{\rm nm}$ and
$l_2=62\,{\rm nm}$ for big and small particles, respectively, compared to
the particle diameters of $4.7\,\mu{\rm m}$ and $2.8\,\mu{\rm m}$
ensure an almost perfect realization of a 2D-system.
The fact that the centers 
of masses of the different particles do not lie on the same plane
can be neglected.
Furthermore, the flatness of the
interface can be controlled in the range of less than $1\,{\mu}{\rm m}$.
Inclination control of the whole setup guarantees a nearly horizontal
alignment of the flat surface, ruling out the occurrence of any density
gradients in the sample.
We control the interaction strengths $\Gamma_{ij}$ between the dipoles
by applying an external magnetic field perpendicular to the
interface. The dipole
interaction dominates all other interactions in this colloidal
system \cite{Maret_1}. The sample was conserved for months and
measurements were taken for a duration of up to
24 hours.
All necessary data were recorded by video microscopy.
Typically, about $1300$ particles
were observed in a box of $689\,\mu{\rm m} \times 505\,\mu{\rm m}$,
with a total
amount of $10^{6}$ particles in the whole sample.
The density ratio $\rho_2/\rho_1$ was varied between $0.67$ and $1$.
Local and statistical properties of the sample are gathered at a
rate of about 1 frame/second on all relevant time and length scales.

\begin{figure}[t]
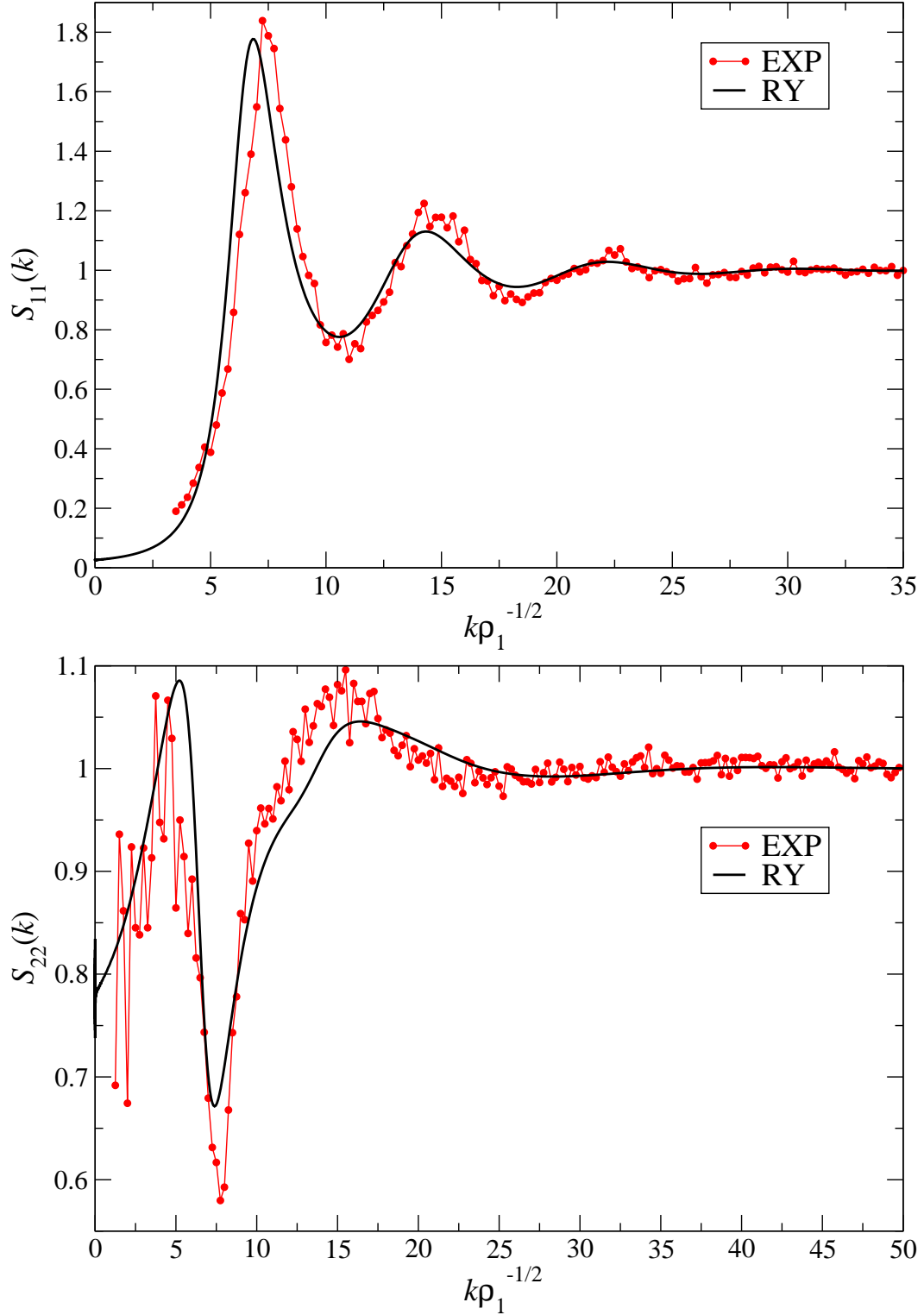

  \begin{center}
    \includegraphics[width=14.0cm,clip=true]
    {./fig3a.eps}
    \includegraphics[width=14.0cm,clip=true]
    {./fig3b.eps}
    \caption{Partial structure factors $S_{11}(k)$ for the big
    particles and $S_{22}(k)$ for the small particles. Experimental
    results (EXP)
    are compared to theoretical ones (RY) for the parameters:
    $\Gamma_{11}=4.05$, $\rho_2/\rho_1=0.89$, $\chi_2/\chi_1=0.1$.}
  \label{fig:sofk_exp_theory}
  \end{center}
\end{figure}

A comparison between the theoretical and experimental structure factors
is shown in Fig.\ \ref{fig:sofk_exp_theory}, pertaining to 
susceptibility ratio $\chi_2/\chi_1 = 0.1$ and corresponding to the
real experimental situation. The representative results are shown
here for density ratio $\rho_2/\rho_1 = 0.89$ and coupling constant
$\Gamma_{11} = 4.05$. Very good agreement between theory and experiment
is achieved, supporting the modeling of the system by means of
dipolar interactions exclusively; evidently, all other residual 
forces in the colloidal suspension are much weaker and can be neglected.
A prepeak is seen in the small-particles structure factor
$S_{22}(k)$, whose height is comparable with that of the second peak
due to the small value of $\chi_2$. Indeed, as the second peak arises
from pure small-small interactions, its height is suppressed since
the latter scale as $\chi_2^2$. The prepeak would trivially
disappear in both limits $\chi_2/\chi_1 \to 0$ (ideal small particles)
and $\chi_2/\chi_1 \to 1$ (one-component system). Again,
$S_{11}(k)$ is deprived of a prepeak.

\begin{figure}[t]
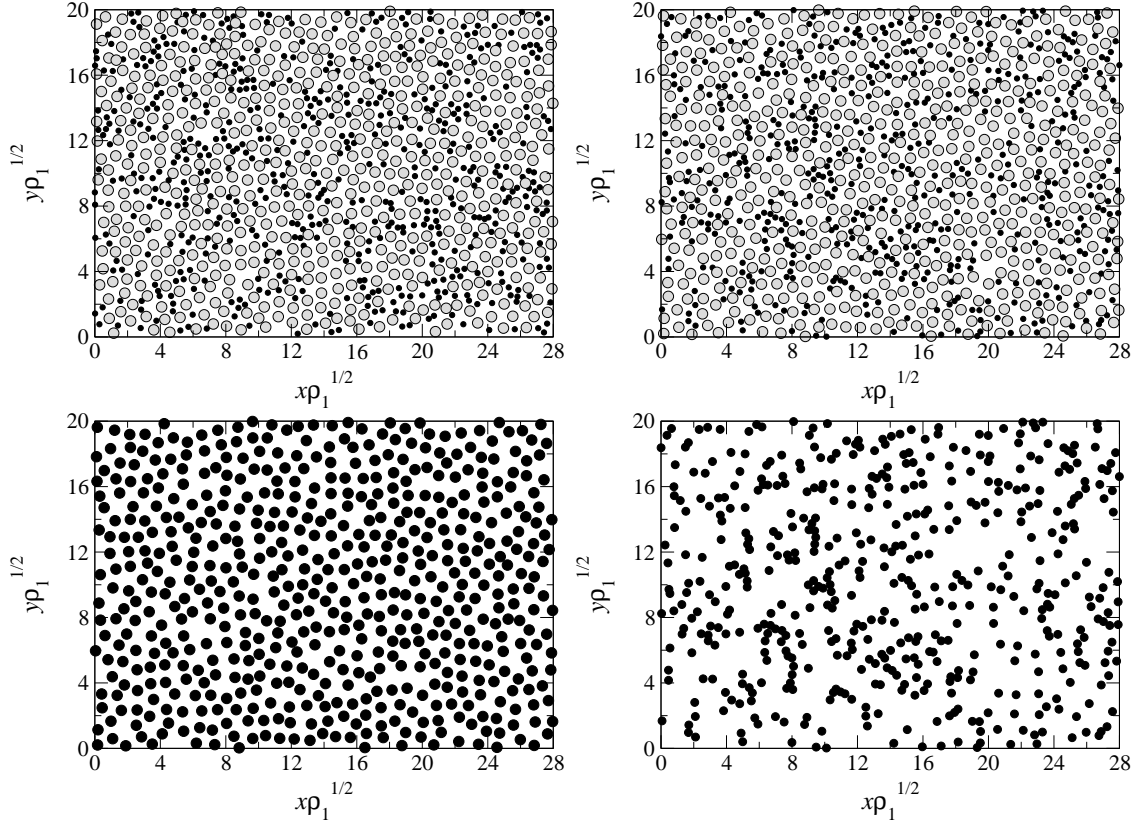

\begin{minipage}{15.0cm}
  \begin{center}
    \includegraphics[width=7.4cm,clip=true]
    {./fig4a.eps}
    \includegraphics[width=7.4cm,clip=true]
    {./fig4b.eps}
  \end{center}
\end{minipage}
\begin{minipage}{15.0cm}
  \begin{center}
    \includegraphics[width=7.4cm,clip=true]
    {./fig4c.eps}
    \includegraphics[width=7.4cm,clip=true]
    {./fig4d.eps}
   \end{center}
\end{minipage}
    \caption{Snapshots of the binary magnetic mixture from experiment
    and simulation for the parameter combination 
    $\Gamma_{11}=4.05$, $\rho_2/\rho_1=0.89$,
    $\chi_2/\chi_1=0.1$. The big particles are denoted gray and
    the small ones black.
    Clockwise from the upper left panel:
    experiment, simulation (both species), simulation
    (small particles only) and simulation (big particles only).} 
  \label{fig:snapshots}
\end{figure}

\begin{figure}[t]
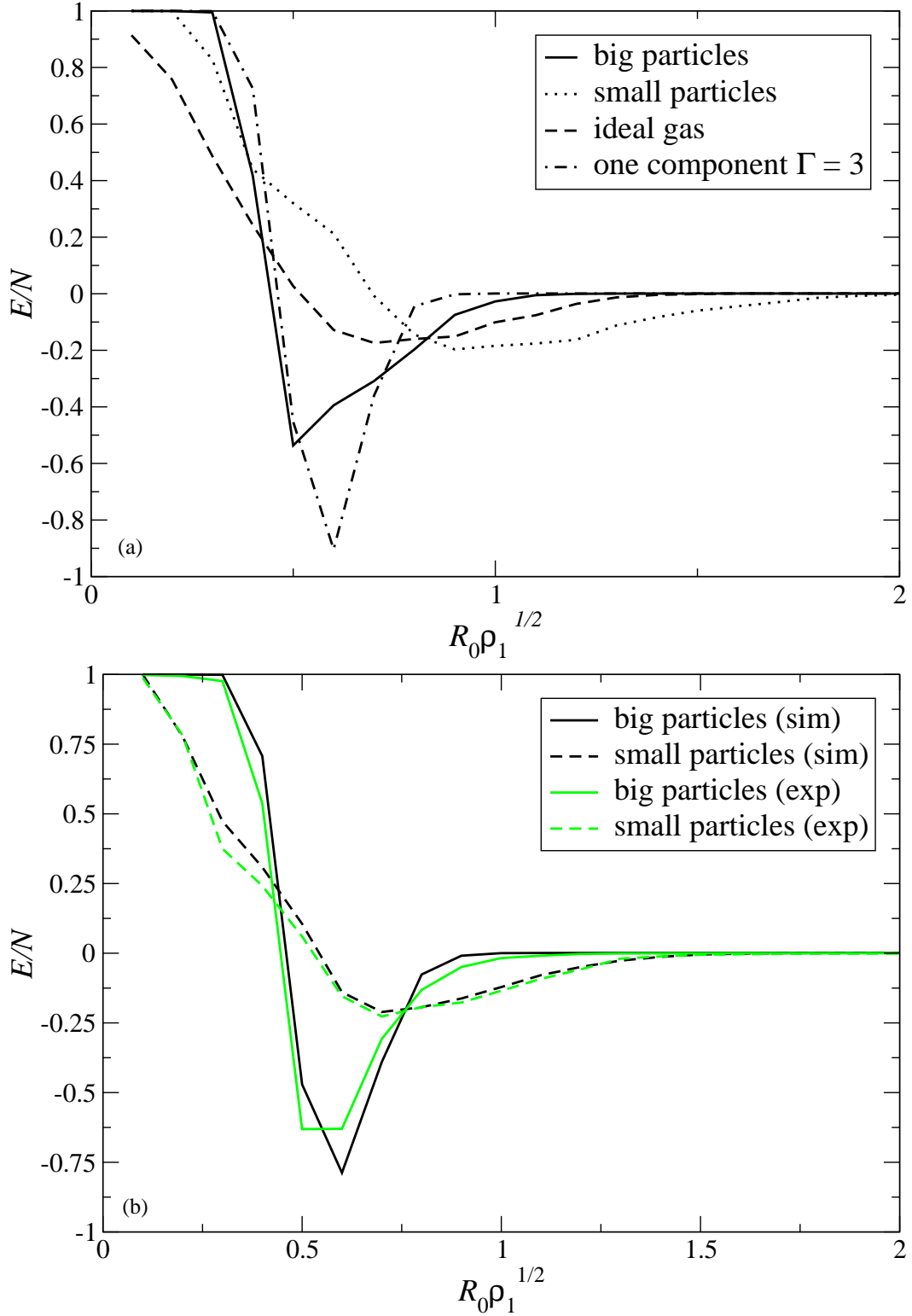

  \begin{center}
    \includegraphics[width=14.0cm,clip=true]
    {./fig5a.eps}
    \includegraphics[width=14.0cm,clip=true]
    {./fig5b.eps}
    \caption{(a) Euler characteristic per particle, $E/N$, 
    from simulation for a binary magnetic
    mixture and
    $\Gamma_{11}=3.0$, $\rho_2/\rho_1 = 0.5$, $\chi_2/\chi_1=0.5$; 
    (b) Comparison of the theoretical (black lines) and experimental
    (gray lines) of the same quantity for 
    the parameters
    $\Gamma_{11}=4.05$, $\rho_2/\rho_1 = 0.89$, $\chi_2/\chi_1=0.1$. 
    Solid lines: big particles;
    dashed lines: small particles.}
  \label{fig:euler}
  \end{center}
\end{figure}

Further evidence of the subclustering phenomenon is provided by
real-space data.
In Fig.\ \ref{fig:snapshots} we show experimental and simulation snapshots of
the system under consideration.
We observe a cluster structure of
the smaller particles that perfectly fits in the voids shaped by the
big particles. To better quantify the morphology of the thus forming
spatial patterns, we employ tools from integral geometry, and in
particular the Euler characteristic $E$, which describes
the topology of the pattern \cite{likos}.
We have discretized the snapshot by superimposing a 
grid on the picture, surround each particle by  circle and
gradually increase the (cover) radius $R_0$ of the latter. 
The Euler characteristic was then calculated on the
discrete lattice \cite{likos}. The
resulting spatial pattern consists then of a number $C$ of connected
components and a number $H$ of holes and
$E = C - H$ \cite{likos}.

In Fig.\ \ref{fig:euler}(a), we show the $E$-curves 
of the two components of a binary mixture, compared to a
one-component and an ideal gas system. The underlying snapshots are
taken from simulation data.  
The Euler characteristic of the
small component tends to zero much more slowly than the ideal gas
curve. For the ideal gas, the particles are more or less evenly spread
over the area, so the vacant space between the particles can be filled
much faster than in the two-component case where the inter-aggregate
space between the small particles is larger.
A comparison between the big component and the one-component system
reveals that the Euler characteristic of the former becomes less
negative and tends to zero much more slowly due to the less and larger
voids caused by the aggregation of the small component.
Fig.\ \ref{fig:euler}(b) refers to the parameter combination of the
experimental sample. There, the $E$ vs.\ $R_0$-curves from 
simulation and experiment are compared, showing very good agreement.
The prepeak position in Fig.\ \ref{fig:sofk_exp_theory}(b) points
to a typical cluster size $L_{\rm cl}/\sqrt{\rho_1} \cong 1.26$
which agrees well with the position of the most negative value of
$E_2$ from Fig.\ \ref{fig:euler}(b) and is consistent with the large
holes left there when the covering circle radius $R_0 \cong L_{\rm cl}/2$. 
The partial clustering is caused by the interplay between the 
successively stronger repulsive interactions $v_{22}$, $v_{12}$ and
$v_{11}$, which preclude macrophase separation and favor the
nesting of small particles in the voids of the big ones.

We have studied equilibrium cluster formation in a 
two-component
colloidal model system. We find partial clustering: only the
small
particles show a clustering, as witnessed by a first peak in their
structure factor while the big particles do not exhibit any peak
on this scale. The cluster structure does not consist
of well-separated isolated clusters 
with a well-defined geometrical shape, such as circular 
or lamellar, which form for interactions with a short-range
attraction and a long-range repulsion \cite{Imperio}. Rather, it
possesses a percolating 
sponge-like topology,
as signaled by a considerably
negative $E$-value for large covering circles. 
We presented results for $\chi_2/\chi_1 = 0.5$
and $0.1$ but the subclustering phenomenon holds for a broad range of
parameter combinations; more extensive results will be published
in the future.

The subclustering behavior is more
general and not just limited to paramagnetic
colloids. Similar effects are encountered in 
binary star polymer mixtures \cite{mayer:pre:04}.
Our experimentally realizable system offers
a clean and reproducible example of this phenomenon which has not
received due attention in two spatial dimensions up to now.
The common feature of systems exhibiting partial
clustering is that they possess a negative non-additivity, 
i.e., if their cross-interaction is less repulsive than the sum of the two
direct interactions.
Partial clustering, however, is absent in additive hard sphere mixtures
\cite{Ashcroft}. In systems with positive non-additivity like
the Asakura-Oosawa-Vrij model \cite{Lekkerkerker}, partial
clustering was never observed, due to the attractions induced by the 
polymers, which drive
macro-phase separation. 
It would be interesting to study the viscosity and the
propagation of light in partial clustered structures in
order to extract their specific material properties. It is further 
tempting
to shock-freeze the clustered structure and 
use it as a bicontinuous device
of controlled, random nano-porosity.

We thank Dominique Levesque 
for helpful discussions, Christian Mayer for providing unpublished
data and Roland Hund for providing raw data of the experiments. 
This work
has been supported by the DFG within the SFB-TR6, Project Sections C2 and  C3.


\begin{thebibliography}{99}

\bibitem{Pine} V. N. Manoharan \textit{ et al.}, Science {\bf 301}, 483 (2003).

\bibitem{Kecht} J. Kecht {\it et al.}, Langmuir {\bf 20}, 5271 (2004).

\bibitem{Lutz} C. Lutz {\it et al.}, Phys. Rev. Lett. {\bf
  93}, 026001 (2004).  

\bibitem{Goedel} F. Yan and W. A. Goedel, Chem. Mater. {\bf 16},
  1622 (2004), Nano Lett. {\bf 4}, 1193 (2004). 

\bibitem{Pusey} P. N. Pusey, in {\it Les Houches, Session} LI, {\it
  Liquids, Freezing and Glass Transition} (North-Holland, Amsterdam,
  1991), ed. by J.-P. Hansen, D. Levesque and J. Zinn-Justin.

\bibitem{Sear} R. P. Sear and W. M. Gelbart, J. Chem. Phys. {\bf 110},
  4582 (1999).

\bibitem{Imperio} A. Imperio and L. Reatto, J. Phys.: Condens. Matter
  \textbf{16}, S3769 (2004).

\bibitem{Malescio} G. Malescio and G. Pellicane, Nature Materials
  \textbf{2}, 97 (2003).

\bibitem{mladek:prl:06} B. M. Mladek {\it et al.}, Phys. Rev. Lett. {\bf 96},
045701 (2006).

\bibitem{sciortino_mossa} F. Sciortino {\it et al.},
  Phys. Rev. Lett. {\bf 93}, 055701 (2004).

\bibitem{mossa_co} S. Mossa {\it et al.}, Langmuir {\bf 20}, 10756 (2004).

\bibitem{Stradner} A. Stradner {\it et al.}, Nature {\bf 432}, 492
  (2004).

\bibitem{Campbell} A. I. Campbell {\it et al.}, Phys. Rev. Lett. {\bf
  94}, 208301 (2005).

\bibitem{Weitz} P. J. Lu {\it et al.}, Phys. Rev. Lett. {\bf 96},
  028306 (2006).

\bibitem{Maret_1} K. Zahn {\it et al.}, Phys. Rev. Lett. {\bf 79}, 175
(1997).

\bibitem{Maret_2} K. Zahn {\it et al.}, Phys. Rev. Lett. {\bf 82}, 2721
(1999); Phys. Rev. Lett. {\bf 85}, 3656 (2000).

\bibitem{Maret_3} C. Eisenmann {\it et al.}, 
Phys. Rev. Lett. {\bf 93}, 105702 (2004).

\bibitem{Naegele} M. Kollmann {\it et al.}, Europhys. Lett. {\bf 58},
  919 (2002). 

\bibitem{Koenig} H. K{\"o}nig {\it et al.}, Eur. Phys. J. E {\bf 18}, 287 (2005).

\bibitem{roth:evans:epl:01} R. Roth and R. Evans, Europhys. Lett. {\bf 53},
271 (2001).

\bibitem{Klein} J. M. M\'endez-Alcaraz {\it et al.}, Physica A {\bf 220},
  173 (1995). 


\bibitem{cnl:epl:05} E. Stiakakis {\it et al.}, Europhys. Lett. {\bf 72},
664 (2005).

\bibitem{allen} M. P. Allen and D. J. Tildesley, {\it Computer Simulation of
  Liquids} (Clarendon Press, Oxford, 1987).

\bibitem{Hansen} J.-P. Hansen and I. R. McDonald, {\it Theory of
  Simple Liquids} (Academic, London, 1986), 2nd ed. 

\bibitem{Rogers} F. J. Rogers and D. A. Young, Phys. Rev. A {\bf 30},
  999 (1984).

\bibitem{DYNABEADS} DYNABEADS M-450, uncoated; Deutsche Dynal GmbH, Postfach
111965, D-20419 Hamburg, Germany.

\bibitem{likos} C. N. Likos {\it et al.}, J. Chem. Phys. {\bf 102}, 9350 (1995).



\bibitem{mayer:pre:04} C. Mayer {\it et al.}, unpublished.

\bibitem{Ashcroft} N. W. Ashcroft and D. C. Langreth, Phys. Rev. {\bf
  156}, 685 (1967).

\bibitem{Lekkerkerker} H. N. W. Lekkerkerker {\it et al.}, Europhys. Lett. {\bf
  20}, 559 (1992).


\end{thebibliography}
\end{document}